%% file: main.tex
\documentclass[sigconf,10pt]{acmart}

\usepackage{shortcuts}
\usepackage{algorithmic}
\usepackage{graphicx}
\usepackage{textcomp}
\usepackage{xcolor}
\usepackage{comment}
\usepackage{subfigure}

\usepackage{tikz}
\usepackage{amsmath}

\usepackage{filecontents}

\copyrightyear{2024}
\acmYear{2024}
\setcopyright{rightsretained}
\acmConference[ACM MobiCom '24]{The 30th Annual International Conference on Mobile Computing and Networking}{November 18--22, 2024}{Washington D.C., DC, USA}
\acmBooktitle{The 30th Annual International Conference on Mobile Computing and Networking (ACM MobiCom '24), November 18--22, 2024, Washington D.C., DC, USA}
\acmDOI{10.1145/3636534.3690709}
\acmISBN{979-8-4007-0489-5/24/11}

\begin{document}
\pagenumbering{gobble}
\settopmatter{printfolios=true}
\date{}

\title{Enabling Physical Localization of Uncooperative Cellular Devices}
\author{Taekkyung Oh$^1$, Sangwook Bae$^2$, Junho Ahn$^1$, Yonghwa Lee$^3$, Tuan Dinh Hoang$^1$, Min Suk Kang$^1$, Nils Ole Tippenhauer$^4$, Yongdae Kim$^1$}
\affiliation{\institution{$^1$KAIST, $^2$Cape, $^3$Theori, $^4$CISPA Helmholtz Center for Information Security}
\country{}}
\renewcommand{\shortauthors}{Oh et al.}

\input{secs/0_Abstract}

\ccsdesc[500]{Networks~Mobile and wireless security}
\keywords{Cellular localization; LTE; Law enforcement}

\maketitle
\widowpenalty 0
\clubpenalty 0
\input{secs/1_Intro}

\input{secs/2_Background}
\input{secs/3_UMA}

\input{secs/4_Challenge}

\input{secs/6_Tracking}
\input{secs/7_Holding}
\input{secs/8_Shadow}

\input{secs/9_Repeater}
\input{secs/10_E2E}

\input{secs/10.5_Feasibility}
\input{secs/12_Discussion}
\input{secs/11_Related}
\input{secs/13_Conclusion}

\bibliographystyle{ACM-Reference-Format}
\bibliography{ref, conf}

\clearpage

\end{document}

%% file: secs/0_Abstract.tex
\begin{abstract}
In cellular networks, authorities may need to physically locate user devices to track criminals or illegal equipment. This process involves authorized agents tracing devices by monitoring uplink signals with cellular operator assistance. However, tracking uncooperative uplink signal sources remains challenging, even for operators and authorities. Three key challenges persist for fine-grained localization: i) devices must generate sufficient, consistent uplink traffic over time, ii) target devices may transmit uplink signals at very low power, and iii) signals from cellular repeaters may hinder localization of the target device. While these challenges pose significant practical obstacles to localization, they have been largely overlooked in existing research.

This work examines the impact of these real-world challenges on cellular localization and introduces the Uncooperative Multiangulation Attack (\umaplus) to address them. \umaplus can 1) force a target device to transmit traffic continuously, 2) boost the target's signal strength to maximum levels, and 3) uniquely differentiate between signals from the target and repeaters. Importantly, \umaplus operates without requiring privileged access to cellular operators or user devices, making it applicable to any LTE network. Our evaluations demonstrate that \umaplus effectively overcomes practical challenges in physical localization when devices are uncooperative.
\vspace{-2mm}
\end{abstract}

%% file: secs/1_Intro.tex
\section{Introduction}
\label{sec:intro}

The broadcast nature of wireless signals allows seamless communication without revealing precise locations. However, this feature challenges authorities who need to pinpoint specific user devices for legitimate reasons. 
For example, the perpetrators engaged in criminal 
 activities~\cite{reaves2015boxed, sahin2017sok, scam_telecom, scam_china} can operate their cellular devices (\eg, SIM boxes) in densely populated urban environments, making it difficult for law enforcement agencies to track and apprehend them.

Authorities may seek assistance from mobile network operators (MNOs) in the process of cellular device localization, and operators are generally willing to cooperate if the requests are valid and comply with legal procedures~\cite{authority_carrier}. 
However, MNOs' assistance in localization comes with certain limitations. 
While they can provide authorities with readily-available, high-level user data, such as serving cell ID or triangulation based on signal strength, the level of precision achieved remains relatively coarse-grained. 
For example, Cell ID (CID) provides only limited positioning of devices with about 1km accuracy~\cite{cherian2013lte, del2017survey, CID_performance}. 
While existing native 3GPP positioning features, such as Observed Time Difference of Arrival (OTDoA), Uplink Time Difference of Arrival (UTDoA), Enhanced CID (E-CID), have been proposed to improve cellular localization, they are all optional features~\cite{3gpp_36305} and thus cannot provide universal fine-grained localization across all MNOs, radio-access network (RAN) vendors, and User Equipment (UE) manufacturers.

Fine-grained LTE localization requires \textit{active, real-time monitoring of uplink signals from targeted UEs}. This necessitates advanced physical-layer approaches beyond conventional control-plane data collection in mobile networks. For instance, Timing Advance (TA) commands, which compensate for signal propagation delays between base stations and UEs, can be used to infer sub-cell-level UE locations~\cite{tan2021data}. Localizers may also employ Time of Arrival (ToA) measurements to determine a target user's distance. LTrack~\cite{kotuliak2022ltrack} leverages both ToA of UE uplink signals and TA commands, and LTEye~\cite{kumar2014lte} measures the Angle of Arrival (AoA) of UE uplink signals using Synthetic Aperture Radar (SAR).

While existing physical-layer approaches show promise, their effectiveness for reliable fine-grained localization of uncooperative devices -- \textit{those that are neither directly controlled by the localizer nor report their own location} -- in real-world scenarios remains unproven. The gap between their performance in controlled environments and our goal of localizing uncooperative devices in realistic scenarios poses significant challenges. This work aims to a) identify real-world obstacles overlooked in cellular localization, b) develop universally applicable solutions, and c) demonstrate a comprehensive process for fine-grained localization of uncooperative cellular devices.

In this paper, we conduct real-world experiments to investigate challenges in applying existing physical localization techniques to uplink signals of uncooperative devices. Our experiments reveal three key issues: \textbf{C1}) Insufficient uplink traffic: Localization becomes impossible if a target UE is not actively transmitting uplink traffic during tracking.  
\textbf{C2}) Low signal power: When a target is close to a cell tower, its uplink signal power may be significantly reduced, hindering signal detection and location determination, especially in urban areas with densely deployed base stations. \textbf{C3}) Interference from repeaters: Cellular repeaters, ubiquitous in modern indoor environments, amplify and relay cellular signals, introducing severe noise in determining a target's location. 

It's important to note that cellular operators' assistance alone is insufficient to address these challenges. While operators have full control over their RAN infrastructure, including base stations (eNBs) and repeaters, and RAN vendors offer various management features via proprietary interfaces, these capabilities do not extend to making arbitrary adjustments to specific UEs' uplink signals. To the best of our knowledge, no existing RAN systems allow operators to finely tune uplink scheduling or power control of a particular UE, which are necessary for achieving reliable, fine-grained cellular localization.

\nibf{Real-world case: Vishing investigation.} 
As a realistic example, police agents in East Asia trace the location of the cellular devices engaged in vishing fraud called SIM boxes~\cite{gsma_vishing,gsma_vishing2,sahin2017sok}, and subsequently confiscate them~\cite{vishing_korea}. 
Note that those SIM Boxes are uncooperative devices. They don't have GPS sensors and only have call-relaying capability. 
The agents are initially provided with SIM boxes' coarse-grained location (\ie, cell ID) from MNOs. They then employ a portable sniffer to track uplink signals emitted by SIM boxes. In this situation, they may encounter the aforementioned challenges, especially in urban cellular networks, and cellular operator's assistance is insufficient to address them.

To overcome these hurdles, we design an Uncooperative Multiangulation Attack~(\umaplus) and show that overcoming these challenges is indeed feasible with \umaplus. 
We reveals that a localizer with \umaplus, after obtaining coarse-grained location information of a target UE from the MNO, can further obtain UE's strong uplink signals reliably. 
We demonstrate end-to-end that real-time manipulation of downlink and uplink signals is feasible, enabling physical localization of uncooperative devices with current, limited operator support. As \umaplus functions in any LTE network adhering to mandatory standard features, it achieves not only reliable but also universal cellular localization. 

\umaplus involves two main approaches: 

\noindent (1) We suggest a \textit{scheduling manipulation} to effectively address \textbf{C1}. With scheduling manipulation, the localizer (e.g. law enforcement), impersonates the target UE and informs the eNB that it has data to transmit. Accordingly, the eNB continuously allocates uplink resources to the target and the target's connection remains active. 
Due to constant resource allocation, the target UE continues to transmit uplink traffic even when it has no data to transmit in its current buffer. 
This scheduling manipulation ensures that the UE is always transmitting, preventing it from becoming idle.

\noindent (2) We propose a \textit{power boosting}, designed to resolve \textbf{C2} and \textbf{C3}. Using power boosting, the localizer, now, impersonates the eNB and instructs the target UE to increase its transmission power to the maximum level. 
The power boosting directly addresses \textbf{C2}, as it allows the localizer to increase the target's signal to its maximum level.
Moreover, the power boosting enables the localizer to overcome the impact of a repeater. 
Since the repeater consistently transmits with constant signal strength, executing the power boosting allows the localizer to observe a notable disparity in signal strength between the target UE and the repeater. 

To sum, \umaplus has three unique characteristics:
\begin{itemize}[leftmargin=*,noitemsep,nolistsep]    
    \item {\bf Network-wide.} Unlike existing techniques, which focus on locating uplink signal sources through physical-layer measurements (e.g., AoA, ToA, TA), \umaplus addresses yet-overlooked systematic challenges that arise when operating localization techniques without device cooperation.
    \item {\bf Universal.} While existing native 3GPP positioning depends on optional features in eNB/UE implementations~\cite{3gpp_36305}, \umaplus operates effectively in virtually any LTE networks that adhere to the core, mandatory standard features. 
    \item {\bf End-to-end.} \umaplus provides publicly available end-to-end demonstration (\ie, from a target's phone number to its physical location) of fine-grained cellular localization; see our public website for its demonstration videos~\cite{uma-site}.
\end{itemize}

Note that localization techniques are often regarded as privacy attacks as they could violate user privacy depending on the context in which the techniques are used. 
While we focus on scenarios that assume authorities with legitimate purposes as main performers of localization techniques, we acknowledge that unprivileged third parties may use \umaplus for unlawful purposes. 
To simplify writing and our arguments,  
in this paper, we refer to the \umaplus technique as attacks. 

\nibf{Responsible disclosure.}
We have reported our findings 
to GSMA. They have acknowledged our contributions with \textit{CVD-2023-0070} and \textit{CVD-2023-0077}.

%% file: secs/2_Background.tex
\section{Preliminaries}

\nibf{Identifiers in LTE.}
When a UE communicates with LTE networks, several identifiers are involved, which are assigned by different network entities. The identifiers can be classified as permanent and temporary. Representatively, the International Mobile Subscriber Identity (IMSI) is a permanent identifier that remains unchanged once assigned to the user's Universal Subscriber Identity Module (USIM).

During radio communication, two temporary identifiers -- Temporary Mobile Subscriber Identity (TMSI) and Radio Network Temporary Identifier (RNTI) -- are mainly used to avoid privacy issues~\cite{imsi_attack_1, imsi_attack_2} caused by the permanent identifiers. TMSI is assigned by core networks when a UE accesses LTE cellular networks and a connection is established.  
RNTI is assigned by an eNB for physical layer communication in cellular radio networks when the UE requests a connection to radio access networks~\cite{3gpp_36321}.
RNTI changes occur when a UE -- eNB radio connection is re-established. If no data transmission occurs for a specified period, the eNB terminates the connection. A new RNTI is then allocated to the UE for its next data transmission. 

\nibf{Downlink Control Information (DCI).}
The eNB and UE transmit their traffic through downlink and uplink channels in the physical layer, respectively. In the physical layer channel, the subframe is a fundamental resource unit for radio communication. 
In order to decode downlink data and enable the transmission of uplink data, 
DCI plays a crucial role, especially in notifying scheduling information and transmission power control for the UEs. Each UE utilizes the RNTI to decode the DCI message and identify the allocated resources. 
Specifically, a certain format of DCI, DCI format 0 (DCI 0), is used for controlling uplink transmission power to ensure optimal signal quality and identifying allocated resources (\ie, uplink grant) on Physical Uplink Shared Channel (PUSCH).  
Note that DCI 0 is broadcast without encryption, it is possible for a third party to obtain a specific UE's uplink scheduling information by using the UE's RNTI.

%% file: secs/3_UMA.tex
\section{Attack Model}
\label{s:untouched}

\subsection{Motivation}
\label{ss:motivation}
Fine-grained localization of cellular devices beyond cell-level precision is highly desired for various applications, including emergency response, criminal investigation, and potentially unlawful surveillance. Localizers typically start with a device's coarse-grained location before attempting to narrow it down further.

However, studies indicate that even authorized agents with mobile operator assistance struggle to achieve fine-grained cellular localization. While authorities can obtain approximate location from operators, this information alone is insufficient for localizing cellular devices ``down to the front door,'' especially when dealing with uncooperative devices. 
A survey from Electronic Frontier Foundation~\cite{eff2021} shows that mobile signal tracking via triangulation by operators offers only about 1~km accuracy.
It also discusses other techniques, such as cell site simulators, which law enforcement agencies use; however, they allow only coarse-grained device presence detection, and work only when the target device supports 2G, a soon-to-be-obsolete technology in the market. 
Peral-Rosado \etal~\cite{del2017survey} also have surveyed localization technologies that can be used by mobile operators. 
It discusses a few technologies that have potential to provide beyond-cell-level localization (\eg, E-CID, OTDoA, and UTDoA) but they are optional features~\cite{3gpp_36305} and thus cannot provide universal fine-grained localization across all MNOs, RAN vendors, and UE manufacturers. As an aside, GPS information can be a useful means to trace the target UE. 
However, authorities cannot access GPS location without the permission of mobile users when they are using iPhone~\cite {iphone_gps} 
and devices employed by criminals may lack support for GPS features.

\label{ss:challenge}
\begin{figure}[!t]
\centering
\includegraphics[width=0.9\columnwidth]{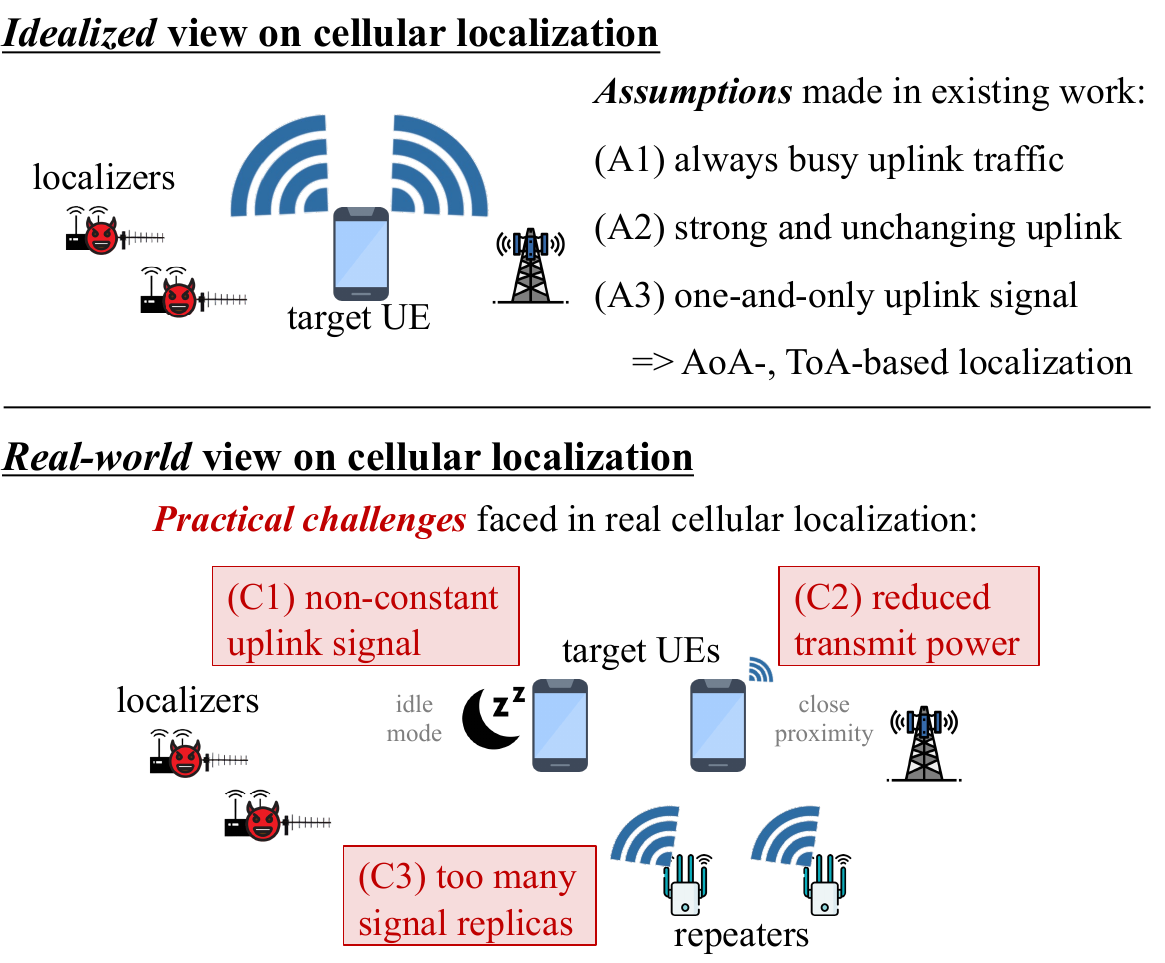}
\vspace{-2mm}
\caption{Idealized view vs. real-world view.}
\label{fig:challenge}
\vspace{-5mm}
\end{figure}
\noindent\textbf{Idealized view on cellular localization.}
There have been several academic studies that attempt to achieve fine-grained localization of cellular devices \textit{outside} the official localization features provided by mobile operators~\cite{kotuliak2022ltrack,kumar2014lte}. 
These studies, however, have rather idealized views on the cellular localization problem, thus overlooking several practical challenges that a localizer may suffer when attempting physical localization in real-world environments.
~\autoref{fig:challenge} illustrates this idealized view on cellular localization used in previous studies.
In this view, a target device is always transmitting uplink signals, which are always strong enough to be detected by the localizer. 
Also, strictly only one uplink signal is transmitted by the target device at a time.  
Thus, the localizer can conveniently conduct physical-layer localization techniques (\eg, AoA, ToA) to locate the target's uplink signal source, after they obtain the target device's temporary network identifier (\eg, RNTI) and detect its uplink signal.

While this simplified view is a good starting point to understand the cellular localization problem, 
in a more realistic, network-wide view on cellular localization, 
the localizer immediately faces several challenges because the real RAN environment is much more complex.  
First, targeted UEs may not be transmitting uplink signals at the time the localizer is attempting to trace them. 
This not only makes localization hard but also makes it difficult to identify the target UE's temporary network identifier. 
Second, target UE's uplink signal power varies significantly because it is always adjusted depending on its channel condition, which may prevent the localizer from detecting low power signals.
Last, the localizer sees not only the target UE's uplink signal but also many replicas of the same signal amplified and relayed by repeaters that are widely deployed to enhance cellular channel quality. 

\subsection{Threat Model}
\label{ss:threat_model}
We consider a scenario where an attacker aims to determine the physical location of target devices that are uncooperative in the localization process, such as not reporting their location or controlling their uplink transmission. In this situation, the attacker cannot directly control the network infrastructure (\eg, eNB or core networks) or run their own code on target devices. This scenario applies to both authorities (\eg, police) and malicious actors. Importantly, this setting does not rely on optional features that are not widely supported in UEs and infrastructure, ensuring broader applicability.

We assume that the attacker knows an online identity~(\eg, phone~number) of the victim and the cell information which the victim is camping on. 
The cell information can be obtained in advance by exploiting the existing presence test~\cite{baektargeted, cheng2023watching, hong2018guti,kune2012location,ludant20225g,shaik2016practical,bitsikas2023freaky} 
or requesting cell information from mobile carriers for investigation~\cite{authority_carrier}. Law enforcement agencies, in particular, can obtain the target's network identifier and cell information with carrier assistance.  Nevertheless, this assistance is insufficient for reliable fine-grained physical localization of uncooperative devices, as mentioned in~\autoref{ss:motivation}. Therefore, in this scenario, carriers are considered semi-cooperative with law enforcement's localization efforts.
 
The attacker employs multiple devices equipped with directional antennas (\ie, AoA) to measure the direction of the target UE's uplink signal. 
Using these devices, the attacker aims to determine the physical location 
of the victim's UE by measuring the direction of uplink signals at multiple points. 
These devices have the capability to passively monitor both uplink and downlink LTE signals. 
This capability can be achieved using an open-source LTE eavesdropper (\eg, LTESniffer~\cite{hoang2023ltesniffer}) with software-defined radio devices~(\eg, USRP~\cite{usrp_x310}). 
One unique capability of attackers is the ability to actively inject malicious LTE signals (\ie, signal overshadowing~\cite{erni2022adaptover,yang2019hiding}) on both downlink and uplink channels. 
Upon obtaining the victim's uplink scheduling information using these devices, the attacker can employ a directional antenna to measure the direction of the uplink signal source, detecting the strongest signal power. 
Utilizing AoA measurements from multiple locations, the attacker performs a multiangulation process to ascertain the victim's precise coordinates.

\input{secs/Tables/txpower_distance}
Additionally, among various cellular localization techniques, we use AoA measurements of uplink signals as a fundamental technique. 
Other techniques such as TA, ToA, and TDoA have some limitations in employing fine-grained localization of uncooperative UEs in real-world environments. 
ToA-based cellular localization~\cite{kotuliak2022ltrack} requires an additional process for fingerprinting the victim's device to account for hardware errors. This is essential for accurately calculating the arrival signal timing, to precisely estimate the victim’s location. In addition, TA-based localization~\cite{tan2021data} shows only sub-cell-level localization performance with 78m granularity.

%% file: secs/Tables/txpower_distance.tex
\begin{table}[!t]
\centering
\caption{Uplink Tx power depending on the distance between the eNB and the UE.}
\vspace{-2mm}
\label{tab:tx_pwr_ue_position}
\resizebox{0.9\columnwidth}{!}{
\begin{threeparttable}
\begin{tabular}{lcccccc} 
\toprule
\textbf{Distance (m)}        & \textbf{10} & \textbf{30} & \textbf{50} & \textbf{70} & \textbf{90} & \textbf{110}  \\ 
\midrule
\textbf{Tx PWR (dBm)}        & -7.04       & 0.49         & 5.05        & 7.65         & 7.2              & 7.56          \\
\textbf{RSRP\textbf{~(dBm)}} & -62.37        & -73.43       & -78.91        & -85.36           & -86.45            & -87.6         \\
\bottomrule
\end{tabular}
\begin{tablenotes}
\scriptsize
\item RSRP: Reference Signals Received Power.
\end{tablenotes}
\end{threeparttable}
}
\vspace{-5mm}
\end{table}

%% file: secs/4_Challenge.tex
\subsection{Challenges}
\label{ss:challenges_uncooperative}

Physical localization of uncooperative devices in real-world environments needs to address the following challenges:
\begin{itemize}[leftmargin=*,noitemsep,nolistsep]
    \item[] \textbf{C1)} \textit{The attacker can track the victim only when the victim's uplink traffic is generated}
    \item[] \textbf{C2)} \textit{The attacker is unable to trace the victim when the victim's signal is undetectable}
    \item[] \textbf{C3)} \textit{The presence of a cellular repeater can interfere with the accurate determination of the victim's direction}
\end{itemize}

\nibf{C1: Dependency on victim's uplink signal.}
Recall that the primary property exploited by localization attackers is monitoring uplink signals generated by the victim's UE. From this perspective, the attacker can initiate localization only when the victim generates uplink traffic. In other words, localization is challenging when the victim is silent (\ie, not transmitting any signal). In such instances, the attacker is compelled to passively wait for the victim's uplink transmission, as they lack access to either the eNB or victim UE.

Additionally, it is important to note that RNTI is a temporary identifier that changes frequently (approximately every 15 to 30 seconds~\cite{bae2022watching, kohls2019lost, rupprecht2019breaking}). 
This implies that when a UE is not actively utilizing any services for a specific duration after connection, its radio connection is terminated by the eNB, and the previously assigned RNTI expires. 
If the UE later requests LTE services from the eNB, a new RNTI will be assigned for radio communication. The attacker may face a challenge in identifying the victim's location if their radio connection and RNTI expire during the localization process. This necessitates the attacker to patiently wait for the victim to establish a new radio connection, after which they must track the victim's new RNTI to resume the localization.

\nibf{C2: Undetectable victim's signal.}
The successful measurement of arrival signals at the attacker’s sniffer is crucial for executing the localization attack. If the sniffer fails to detect the victim’s uplink signal, estimating the victim’s location becomes challenging. This issue arises because the UE does not consistently transmit its signal with high power to the eNB. Notably, when the UE is situated in close proximity to the eNB, weak uplink transmission power is used. The UE dynamically adjusts its uplink transmission power to compensate for signal path loss, influenced by factors such as distance and obstacles, in coordination with the eNB~\cite{3gpp_36213,michalevsky2015powerspy}. For instance, \autoref{tab:tx_pwr_ue_position} demonstrates how the UE adjusts its transmission power based on both distance and RSRP.

The transmission power control between the eNB and UE creates a \textit{shadow area} where the victim's signal remains undetectable by the sniffer. To illustrate this shadow area in real-world environments, we examine how the UE's position influences the sniffer's signal detection in an operational commercial network. In our experiment, the sniffer, positioned 80m away from the target UE, successfully detected uplink signals from the UE located 90m away from the eNB. However, when the UE was only 10m away from the eNB, the sniffer failed to detect uplink signals at the same position (refer to our website~\cite{uma-site} for detail). 
Unfortunately, due to the attacker's lack of permission to control networks and UE, they are unable to legitimately instruct the UE to increase its transmission power when encountering the shadow area.

\input{secs/Figures/repeater}

\begin{figure*}[!ht]
\centering
\includegraphics[width=0.8\linewidth]{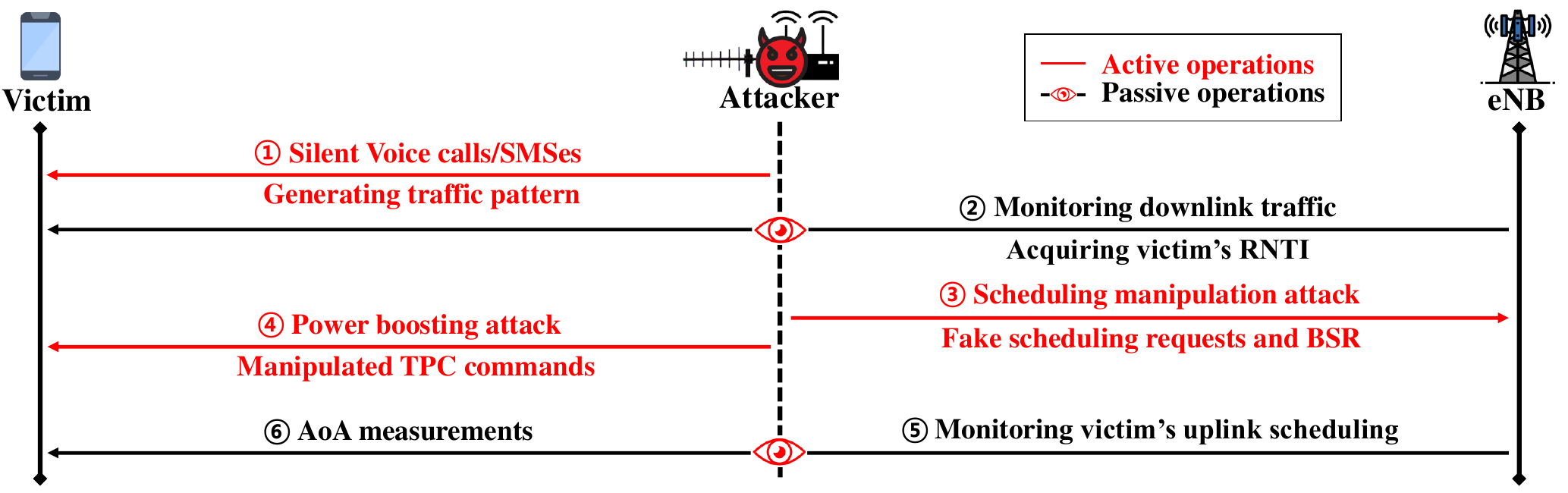}
\vspace{-2.5mm}
\caption{Overview of \umaplus.}
\label{fig:overview}
\vspace{-4mm}
\end{figure*}
\label{ss:uma_overivew}

\nibf{C3: Differentiating victim from repeater.}
A cellular repeater is designed to amplify both downlink and uplink cellular signals, reducing signal path loss between the eNB and UE. 
As a consequence of the repeater's operation, the UE tends to transmit uplink data with lower signal power. 
Concurrently, the repeater transmits (\ie, relays) the data with amplified signal power. 
It is widely deployed to enhance cellular service coverage. 
However, it poses a challenge for attackers in determining the direction of the strongest signal.

We further investigate the influence of cellular repeater on UE's transmission power and sniffer's measurement in an operational commercial network. 
In our experiment, a commercial cellular repeater and a UE are placed inside a building, with the UE positioned 10m away from the repeater and the repeater's external antenna installed outside the window. 
We position a sniffer 30m away from the building and measure RSRP and transmission power at the UE side when the repeater is active and inactive, respectively. 
Subsequently, we measure the arrival signal strength in the direction of the UE and the repeater's external antenna when the repeater is active~(\autoref{fig:repeater}). 
Our results confirm that the repeater's operation increases RSRP (10dBm$\uparrow$) and decreases the UE's transmission power (5dBm$\downarrow$). 
Furthermore, we observe that the arrival signal strength in the direction of the UE is lower than in the repeater's direction. 
These imply that the presence of a repeater can thwart location tracking in environments where cellular repeaters are installed. 

Hence, the attacker needs the capability to discern the direction leading to the UE's location. Despite its substantial impact on localization performance, none of the previous studies have considered the influence of the repeater.

\subsection{Approach: \umaplus}
We present an Uncooperative Multiangulation Attack (\umaplus), which resolves the key challenges of tracking uncooperative devices~(\autoref{ss:challenges_uncooperative}). 
\umaplus leverages AoA measurements with directional antennas, composed of six steps (\autoref{fig:overview}).
These steps include RNTI acquisition using the victim's online identity (\ding{172}--\ding{173}), manipulation of uplink scheduling (\ding{174}), boosting the target's uplink signal strength (\ding{175}), and measurement of the AoA of the uplink signal (\ding{176}--\ding{177}). 
Note that while we assume authorities can obtain the RNTI with assistance from operators, as described in~\autoref{ss:threat_model}, we include the RNTI acquisition process using online identity for the sake of completeness in explaining the overall tracking process in  \umaplus.

In particular, we come up with two approaches to effectively address the challenges of localizing uncooperative cellular devices; the scheduling manipulation attack and the power boosting attack.
The scheduling manipulation attack is specifically designed to address challenge \textbf{C1}, allowing the attacker to acquire the unexpired victim’s RNTI and ensure continuous uplink transmission. 
To overcome challenges \textbf{C2} and \textbf{C3}, we employ the power boosting attack, enabling the attacker to increase the victim’s uplink transmission power and distinguish the victim’s signal from the repeater’s. Both attacks exploit vulnerable protocols in LTE specification, which lack implemented security protection measures.

\nibf{Overall procedures of \umaplus.}
The first two steps are to obtain the victim's RNTI associated with their phone number~(\autoref{s:identify_id}).
This is accomplished by sending silent Short Message Service (SMS) messages (\ding{172}) and monitoring traffic patterns on the downlink channel (\ding{173}). 
Next, the attacker proceeds with a scheduling manipulation attack (\ding{174}); it transmits fake scheduling requests, and reports fake transmission buffer status to the eNB, posing as the victim UE with its RNTI~(\autoref{s:active}). This attack ensures the victim’s radio connection remains active, and their RNTI stays unexpired. Through the scheduling manipulation attack, the attacker forces the victim UE to generate uplink traffic even when there is no data to transmit in its current buffer. 
Subsequently, the attacker conducts a power boosting attack (\ding{175}) by injecting manipulated commands for transmission power control logic into the victim’s UE, disguising as the eNB. 
This attack forces the victim UE to increase its uplink transmission power up to a maximum level (23dBm), overcoming practical challenges caused by the undetectable victim's uplink signal and the presence of a cellular repeater in the radio network~(\autoref{s:tpc}). After these active operations, the attacker obtains the victim’s scheduling information broadcasted over DCI 0 by monitoring the downlink channel (\ding{176}). With this information, the attacker identifies the victim’s signal on the uplink channel. Then, the attacker measures the AoA of the victim's uplink signal by rotating a directional antenna mounted on the sniffer (\ding{177}). 
Ultimately, the attacker can achieve physical localization by tracking down the victim's location iteratively or employing multiangulation with multiple sniffers.

%% file: secs/Figures/repeater.tex
\begin{figure}[t]
\centering
\includegraphics[width=0.78\columnwidth]{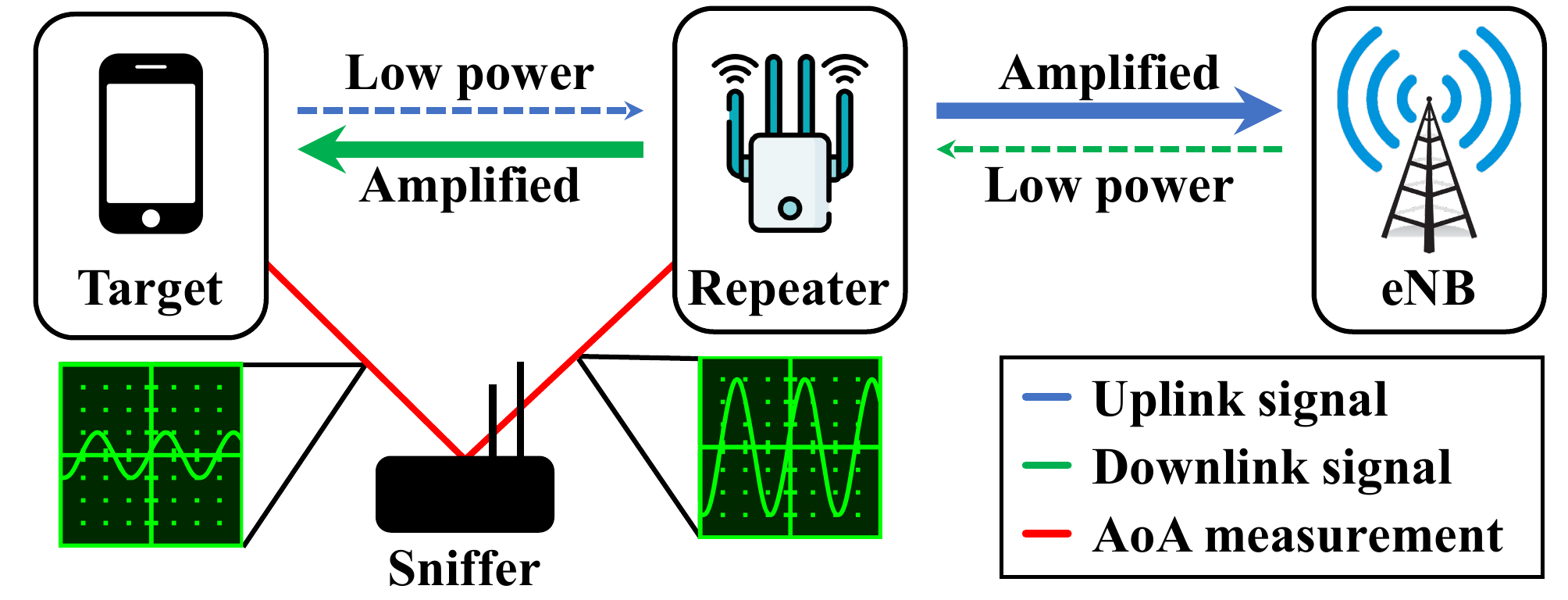}
\vspace{-2mm}
\caption{Impact of cellular repeater in localization.}
\label{fig:repeater}
\vspace{-5mm}
\end{figure}

%% file: secs/6_Tracking.tex
\section{Design of \umaplus}
This section presents the design of our approaches. 
Firstly, we introduce a simple yet feasible method, inspired by previous works, for acquiring the victim's RNTI using only their online identity (\autoref{s:identify_id}). Next, we describe how an unprivileged attacker compels the victim to generate uplink traffic and maintains the victim's RNTI (\autoref{s:active}). Finally, we illustrate how the attacker compels the victim to increase their uplink transmission power and overcomes cellular repeaters (\autoref{s:tpc}).

\subsection{From Online Identity to Radio Identifier}
\label{s:identify_id}
For an attacker armed solely with the target's online identity (\eg phone number, WhatsApp ID), the initial step in \umaplus is to acquire the target's temporary identifier used in the radio network (RNTI). The eNB broadcasts uplink scheduling information to each UE, designating a recipient UE by using the RNTI. Consequently, an unprivileged attacker seeking to monitor and ascertain the uplink signal of the victim UE must be aware of the victim's RNTI. Since the RNTI is intentionally designed to be decoupled from the online identity, acquiring the RNTI becomes essential to establish the link between the victim's online identity and its corresponding RNTI.

Inspired by prior works~\cite{hong2018guti, hussain2019privacy, kohls2019lost, kune2012location, ludant20225g, rupprecht2019breaking, shaik2016practical}, our approach to acquiring the RNTI involves correlating voice/SMS traffic pattern directed toward the victim with monitored downlink packets identified by the recipient's RNTI. 

RNTI acquisition is conducted through the following three steps.
\ding{192} Given the victim's online identity, the attacker successively generates silent voice calls or silent SMSes to the victim following a specific traffic pattern (\eg, multiple times with a constant time gap).
\ding{193} The attacker monitors the downlink traffic of voice calls and SMSes.
Specifically, they monitor the dedicated channels, called bearers, used for sending control messages of calls and SMSes. 
Note that the attacker can monitor the targeted traffic pattern clearly as each bearer is designed to deliver a different type of data. 
For example, messages for controlling voice calls are delivered over a bearer named Data Radio Bearer (DRB) 1, and user-plane traffic and SMS control messages are flooded via DRB 2 and Signalling Radio Bearer (SRB), respectively.  
\ding{194} The attacker examines radio connections and finds out the connection showing the intended traffic pattern. Finally, the attacker determines the RNTI of that connection as belonging to the victim. 

%% file: secs/7_Holding.tex
\subsection{Manipulating~Uplink~Scheduling}
\label{s:active}
One of the key pillars of \umaplus lies in its ability to a) compel the victim to generate uplink transmission and b) retain the acquired victim's radio identifier~(RNTI). To achieve this, we introduce a novel approach known as the \textit{Scheduling Manipulation Attack}. Essentially, the \textit{Scheduling Manipulation Attack} injects a crafted message into the eNB, prompting the eNB to schedule resources for uplink transmission to the victim UE and triggering the UE to transmit uplink data.

\subsubsection{Uplink Scheduling Procedure}
\label{ss:scheduling}
We begin by outlining the control mechanism governing a UE's uplink data transmission. Essentially, the UE is granted permission to transmit uplink data only when it receives uplink grants allocated by the eNB. The UE follows a defined procedure to secure uplink grants for data transmission to the eNB. Initially, the UE transmits a scheduling request~\cite{3gpp_36213}, signaling to the eNB its intention to transmit uplink data. Upon receiving the scheduling request, the eNB grants approval for the UE's Buffer Status Reporting (BSR)~\cite{3gpp_36321}, a mechanism used to report the size of uplink data in its buffer. Subsequently, the UE initiates the BSR, leading to the allocation of sufficient uplink grants by the eNB, allowing the UE to transmit its data.

\input{secs/Figures/scheduling_overview}

\nibf{Scheduling Request} is a PHY layer message sent from the UE to the eNB over the Physical Uplink Control Channel (PUCCH). To transmit a scheduling request to the eNB, the UE uses \textit{schedulingRequestConfig}~\cite{3gpp_36213,3gpp_36331}, which was previously shared via the \tts{RRC connection setup} message. This configuration notifies each UE which resource should be used to request scheduling.

\nibf{Buffer Status Reporting (BSR)} is a MAC layer message sent over PUSCH to notify the eNB of the data awaiting transmission in the UE's buffer. Following the reception of the BSR, the eNB issues multiple uplink grants to the UE, ensuring it has ample resources to transmit the queued data.

\subsubsection{Design of Scheduling Manipulation Attack}
\label{ss:sched_design}
Our approach exploits BSR~\cite{tan2021data} and scheduling requests, which manipulate the victim's uplink scheduling procedure. 
Our work makes use of them to induce unintended uplink transmissions and maintain an unexpired radio connection. 
Scheduling requests and BSR, governed by the PHY and MAC layers, lack security protection such as encryption or integrity checks. 
The core idea is to manipulate uplink scheduling by falsifying the victim's uplink channel, prompting the eNB to allocate uplink grants to the victim, even when the victim has no actual uplink data to transmit. Our approach unfolds through four distinct steps, as illustrated in~\autoref{fig:sched_attack_proc}.

\nibf{1) Acquiring victim's radio configuration.}
The \umaplus attacker seeks to identify the victim's user-dedicated radio configuration, including \textit{schedulingRequestConfig}. This configuration is sent to each UE by the eNB during radio connection establishment via the \tts{RRC Connection Setup} message. As this signaling is unencrypted, the attacker can discern the victim UE's configuration by monitoring the downlink channel when the victim initiates a radio connection with the eNB. From this intercepted data, the attacker can extract the victim's \textit{schedulingRequestConfig} associated with its RNTI.

\nibf{2) Establishing forged uplink channel.}
The \umaplus attacker forges the victim's uplink channel using the acquired radio configuration and RNTI of the victim. Initially, the attacker establishes a radio connection with the eNB to create a communication channel. Subsequently, the attacker configures its RNTI and \textit{schedulingRequestConfig} with identical values to those of the victim. This setting allows the attacker to gain control of the victim's communication channel. Consequently, the attacker impersonates the victim UE, while the victim maintains its legitimate connection.

\nibf{3) Manipulating scheduling request.}
After acquiring the victim's communication channel, the \umaplus attacker injects a fake scheduling request to the eNB over the victim's resource on PUCCH, which is obtained by forging the victim's uplink channel. Since the scheduling request, indicating the UE has data to transmit, is not authenticated, the attacker can forge this request. Subsequently, the eNB allocates uplink resources to the victim's RNTI on PUSCH, informing the victim UE it can send packets and report its buffer status.

\nibf{4) Reporting fake buffer status.}
To obtain multiple grants, the attacker transmits a packet to the eNB over the allocated resource through the fake scheduling request, containing arbitrary data and a fabricated BSR. The fabricated BSR consists of the required buffer size for future transmission and the Logical Channel ID (LCID). In this attack, we set the required buffer size to 200 bytes. The LCID has a value of 3 which indicates data bearer communication in uplink logical channels. Simultaneously, the victim UE transmits a packet with a BSR informing no pending data. Both transmissions occur in the same subframe, but the attacker's signal overshadows the victim's due to the capture effect~\cite{yang2019hiding,erni2022adaptover}. The eNB then issues multiple uplink grants to the victim's RNTI based on the fabricated BSR. In response to these unexpected grants, the victim UE transmits dummy packets containing padding data and a BSR indicating no further data to transmit. 

The attacker repeatedly conducts the above process to have the eNB continuously allocate uplink grants to the victim. During this process, manipulated messages are transmitted only using 10\% of allocated uplink resources to minimize the impact on the victim's Quality of Service (QoS).

%% file: secs/Figures/scheduling_overview.tex
\begin{figure}[!t]
\centering
\includegraphics[width=0.8\columnwidth]{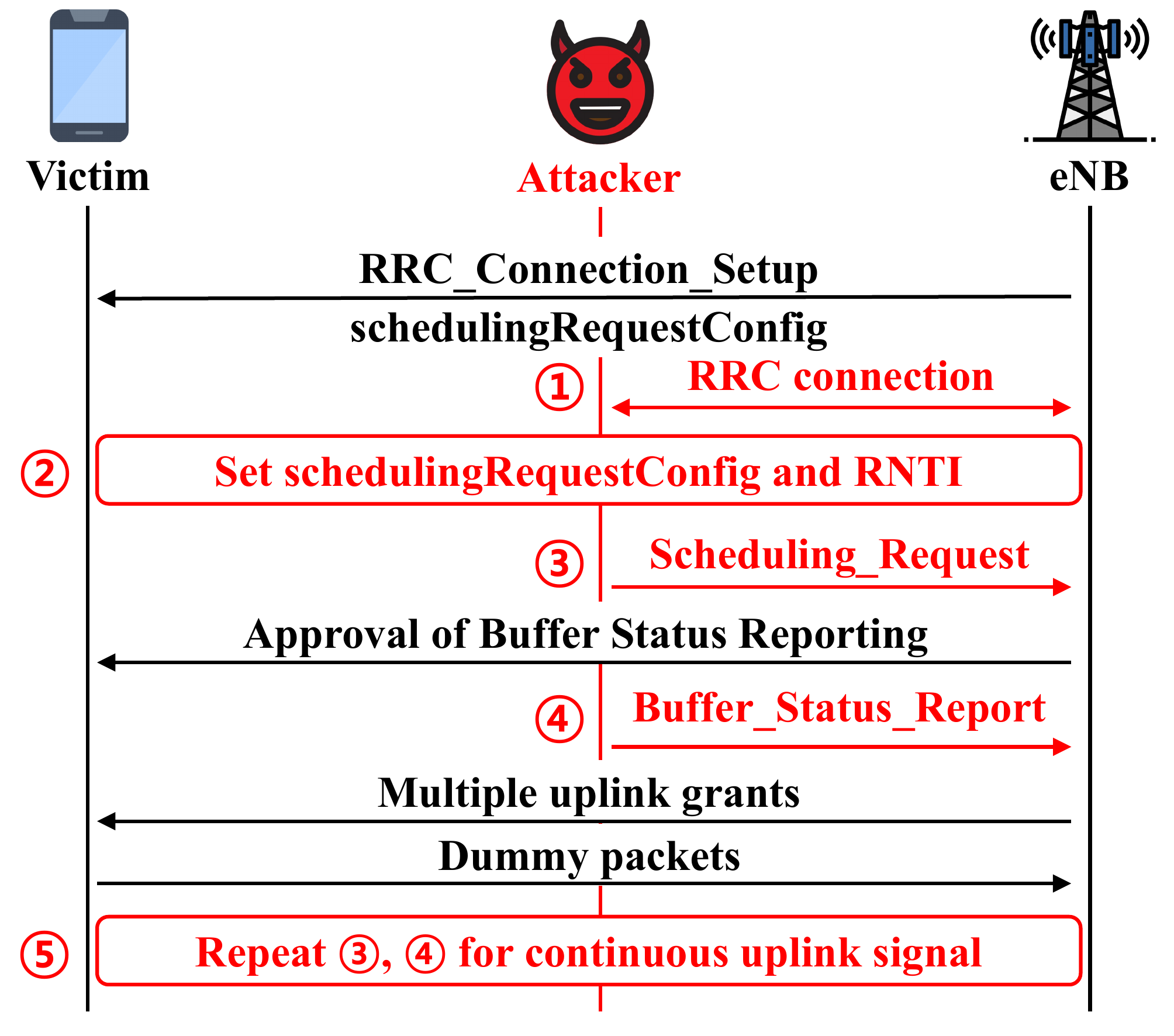}
\vspace{-2mm}
\caption{Procedure of scheduling manipulation attack.}
\label{fig:sched_attack_proc}    
\vspace{-5mm}
\end{figure}

%% file: secs/8_Shadow.tex
\subsection{Boosting Victim's Uplink Signal}
\label{s:tpc}
We present a novel approach called \textit{Power Boosting Attack} to enforce a UE to increase its uplink transmission power. 
This attack is required to overcome a \textit{shadow area} where the victim's signals are undetectable to the localization attacker and the impact of a cellular repeater. 
We first introduce a mechanism for the eNB to control the UE's uplink transmission power~(\autoref{ss:tpc_intro}). 
We then present the implementation of the power boosting attack exploiting the mechanism~(\autoref{ss:power_design}).

\subsubsection{Transmit Power Control}
\label{ss:tpc_intro}

\input{secs/Tables/tpc}

An eNB is responsible for all aspects of data transmission with the connected UEs.
Along with the data transmission scheduling, it also controls the data transmission power of the UE by Transmit Power Control (TPC) command, contained in DCI 0 messages.
The UE adjusts its transmission power based on the internal power control algorithm, which uses a designated value in the TPC command along with the current self-monitored channel quality~\cite{3gpp_36213}. 
One widely used algorithm adopts the \textit{accumulated power control} that gradually adjusts the transmission power according to TPC commands as defined in \autoref{tab:delta_tpc}.

\subsubsection{Design of Power Boosting Attack}
\label{ss:power_design}
Our novel approach to changing the target UE's uplink transmission power without any privilege utilizes TPC command.
TPC commands are delivered via DCI 0 messages without any security protection (neither encryption nor integrity checks).
This leaves them exposed to downlink overshadowing attack~\cite{yang2019hiding}.
The key idea is to inject manipulated subframes containing DCI 0 broadcasts to the victim UE, with TPC set to increase the victim UE's transmission power.
This power boosting attack can be executed by following steps.

\nibf{1) Acquiring configuration for radio communication.}
To identify the structure of a legitimate downlink subframe, the \umaplus attacker retrieves the physical configuration of the target eNB. 
This configuration includes target Physical Cell Identification (PCI), channel bandwidth, Physical channel HybridARQ Indicator Channel (PHICH) configuration, cyclic prefix, and transmission mode of the eNB. 
This can be retrieved by decoding messages that the eNB broadcasts such as Master Information Block (MIB), Primary Synchronization Signal (PSS), and Secondary Synchronization Signal (SSS).

\nibf{2) Manipulating subframe.}
\umaplus attacker crafts a subframe that includes a malicious DCI 0 message using the obtained physical configuration. 
The manipulated DCI 0 messages then consist of 
a) the victim's RNTI, b) arbitrary scheduled uplink transmission information, and c) the TPC command having a value of 3 to boost the uplink transmission power.

\nibf{3) Overshadowing manipulated subframe.}
The \umaplus attacker extends the signal overshadowing (SigOver) attack \cite{yang2019hiding}, injecting a crafted subframe with a proper subframe number over the downlink control channel. More precisely, we mainly modify the subframe construction logic of SigOver \cite{yang2019hiding}, while using nearly the same approach for subframe injection. 
In this work, we inject a manipulated subframe to overshadow every legitimate subframe with index number 9.
This choice serves two purposes: 1) to minimize the impact on the victim's QoS, and 2) to address a race condition with the eNB.
Since the eNB also attempts to adjust the UE's uplink transmission power, the UE decreases its power upon receiving DCI 0 messages from the eNB. 
Consequently, we repeatedly conduct the overshadowing attack with the manipulated subframes.

%% file: secs/Tables/tpc.tex
\begin{table}
\centering
\caption{$\Delta_{PUSCH}$ according to TPC command in DCI 0.}
\vspace{-3mm}
\label{tab:delta_tpc}
\resizebox{0.7\columnwidth}{!}{%
\begin{tabular}{lcccc} 
\toprule
\textbf{TPC command in DCI 0}                                  & 0  & 1 & 2  & 3   \\ 
\midrule
\textbf{Accumulated $\Delta_{PUSCH}$ (dB)} & -1 & 0 & +1 & +3  \\
\bottomrule
\end{tabular}
}
\vspace{-4mm}
\end{table}

%% file: secs/9_Repeater.tex
\subsubsection{Strategy for defeating cellular repeater}
\label{ss:strategy_repeater}
\umaplus could distinguish the victim's signal from the repeater's signal by employing a power boosting attack. 
The key idea is to exploit the cellular repeater's operational logic, specifically designed to amplify the input signal with its maximum output power, enhancing the overall LTE channel quality.
Thus, even after subjecting the victim UE to the power boosting attack, uplink signals amplified by the repeater remain unchanged.

In this context, the attacker conducts the following procedures. 
First, the attacker injects a manipulated TPC command having a value of 3 over DCI 0 message.
Second, both the victim UE and the repeater receive that message.
Third, the victim UE increases its uplink signal strength, but the repeater does not, as it already transmits (relays) the uplink signal with its highest power.
As a result, it leads to an increase in the arrival signal strength only on the victim UE side, as observed from the sniffer's perspective. 
Based on this strategy, the \umaplus attacker can correctly determine the victim UE's direction, overcoming the practical challenge posed by the presence of the cellular repeater.

This strategy is viable due to the distinct operational logic of the UE and repeater. The repeater serves as a passive network component, solely amplifying the input signal with its maximum output power before relaying. 
In contrast, the UE dynamically adjusts its uplink transmission power based on the signal path loss, a topic discussed in~\autoref{ss:challenges_uncooperative}. Consequently, when subjected to the power boosting attack, the UE and repeater exhibit different responses. The repeater consistently amplifies the signal with its maximum power output, while the UE adapts its transmission power dynamically. This discrepancy in reactions to power boosting allows the attacker to distinguish and identify the victim's signal effectively.

%% file: secs/10_E2E.tex
\input{secs/Tables/exp-summary}
\input{secs/Tables/tested_devices}

\section{Evaluation of \umaplus}
\label{ss:real_eval}
We demonstrate the \umaplus in three different testing environments~(\autoref{tab:exp-summary}): a lab testbed (LT), a commercial testbed (CT), and an operating commercial network (ON). 
First, we validate the RNTI acquisition~(\autoref{ss:rnti_eval}) in a busy commercial network, and then prove scheduling manipulation~(\autoref{ss:sched_eval}) and power boosting~(\autoref{ss:power_eval}) in the lab testbed. Subsequently, we demonstrate scheduling manipulation and power boosting attacks in the commercial testbed, highlighting their feasibility~(\autoref{ss:integration}). 
An end-to-end evaluation of \umaplus is then presented, showcasing how the proposed approaches can be integrated to perform the physical localization of uncooperative cellular devices~(\autoref{ss:e2e_multi}). 
Finally, we show the effectiveness of the power boosting attack to distinguish the UE from the cellular repeater in the commercial testbed~(\autoref{ss:strategy_eval}).

\nibf{Experimental environments.}
The lab testbed is built using an open-source LTE platform, srsRAN~\cite{srsran}. 
The commercial testbed, established by a national institute, comprises an industry-grade LTE solution from Nokia, situated in a shielded room measuring 5m $\times$ 7m. 
In addition, we used the operational commercial network of a mobile carrier with the largest subscriber base in South Korea. 
We evaluated scheduling manipulation and power boosting using nine COTS devices in testbed environments.
Both attacks were implemented by modifying srsRAN~\cite{srsran} with USRP X310~\cite{usrp_x310}.

\subsection{RNTI Acquisition}
\label{ss:rnti_eval} 
We first assess its ability to identify the victim's RNTI based on their online identity, in a crowded operational network.

Initially, we investigate the existence of a distinguishable voice/SMS traffic pattern. To do so, we build a dataset comprising over 4 hours of downlink traffic from two operational LTE networks in two countries. This dataset involves 10–60 active users connected to the eNB and a total of 2,142 radio connections. 
We could observe that, as an example case, none of the users has a traffic pattern featuring multiple messages/calls with a time gap of over 6 seconds. It suggests that the attacker could identify the target UE's RNTI by carefully designing a transmission strategy for the distinctive pattern.

With the identified traffic pattern and the target's online identity, we execute the RNTI acquisition process in a crowded commercial network with 862 active users detected over an hour. Using an iPhone 14 Pro (sender) and a Galaxy S10 5G (receiver), we conduct ten trials.
We could confirm that the connection identified by the target's RNTI is distinguishable, achieving a 100\% success rate over all trials.

\subsection{Scheduling Manipulation Attack}
\label{ss:sched_eval}
We evaluate the feasibility of the scheduling manipulation attack. 
For this, we use nine COTS devices in the testbed environment (\autoref{tab:devices}). 
To prevent the UE from requesting uplink grants during the attack, we disable all applications.

First, we assess whether a) the victim’s UE generates uplink traffic even when no data is available for transmission and b) uplink grants are allocated by the eNB.  
To inspect the UE's uplink transmission, we collect uplink packets of the MAC layer at eNB during scheduling manipulation. 
Next, we investigate whether the victim’s RNTI remains unchanged during the manipulation of uplink scheduling and monitor the radio status of the UE on both the srsENB and UE sides.

Surprisingly, we verify that the UE transmits uplink traffic to the eNB whenever uplink grants are allocated, even when these grants are assigned through manipulated scheduling requests and BSR (and it has no data to transmit in its buffer).
Additionally, through the scheduling manipulation attack, the radio connection between the UE and eNB is consistently maintained and the target UE's RNTI remains unchanged. 

Our website~\cite{uma-site} provides packet capture images during scheduling manipulation. These captures illustrate the result of executing the attack four times consecutively, including the contents of both the attacker's and the target's packets.

We further investigated the root causes of continuous uplink transmission even when the UE has no data to transmit in its current buffer. We examine the memory status by dumping the  Communication Processor (CP) called RAMDUMP~\cite{golde2016breaking}, and confirm that uninitialized data (\ie, previously transmitted buffer) in CP memory is transmitted.

\subsection{Power Boosting Attack} 
\label{ss:power_eval}
We investigate whether the TPC injection indeed increases UE's transmission power up to the maximum level.
To demonstrate the effectiveness of the power boosting attack in \umaplus, we run the experiment in the testbed environment, using six COTS devices. 
In this evaluation, we initially execute the scheduling manipulation attack before injecting TPC commands to show the feasibility of an integrated approach. 
Before and after the attack, we measure the transmission power of the UE using a diagnostic monitoring tool, XCAL~\cite{xcal}, to show the increase of uplink transmission power. 

As a result, we confirm that while UE's transmission power is about 10dBm before performing the attack, through the power boosting, the transmission power of UE is increased up to 23dBm (\ie, maximum transmission power). 
Considering the transmission power of the UE presented in \autoref{tab:tx_pwr_ue_position}, it is feasible to overcome the shadow area by increasing the transmission power to 23dBm through power boosting.

\input{secs/Figures/e2e_setup}
\subsection{Demonstration in Commercial Testbed}
\label{ss:integration}
We now move to the commercial LTE testbed and examine the feasibility of the combined attack (scheduling manipulation and power boosting attack). 
We execute the scheduling manipulation attack as the first step to compel the UE to generate uplink transmissions and maintain the UE's RNTI. Subsequently, we conduct the power boosting attack to increase the UE's uplink transmission power while scheduling manipulation is ongoing. 
As a result, we confirm that the combined attack works for three types of devices and commercial network equipment. 
Demonstration videos illustrating the combined attack are available on our website~\cite{uma-site}. 

\subsection{End-to-End Evaluation}
\label{ss:e2e_multi}
Now, we demonstrate an end-to-end evaluation on \umaplus, where an adversary executes four steps in a sequence: 1) RNTI acquisition, 2) Scheduling manipulation, 3) Power boosting, and 4) Physical localization (\ie, multiangulation). 

\nibf{Setup.}
We conduct the experiment in $15m \times 10m$ lab environment as shown in~\autoref{fig:e2e_setup}. 
Note that, due to ethical concerns regarding scheduling manipulation and power boosting attacks, we built the testbed with srsRAN~\cite{srsran}.
Even though we install srsENB using high-performance SDR, USRP X310~\cite{usrp_x310}, its cellular range is inevitably limited. Therefore, we arrange the setup in a spatially constrained room. 
We deploy two sniffers equipped with the directional antenna~\cite{hyper}, one near the eNB and another at the border of the cellular range, respectively. 
As depicted in the figure, the target UE is at nine points. 
For each trial, 
we turn on extra two UEs at the same time, connecting them to the eNB. Afterward, our evaluation is designed to figure out where one of the UEs (\ie, target UE) is physically located. 
Especially, since srsENB doesn't support SMS, we utilize WhatsApp for RNTI acquisition~\cite{shaik2016practical}. 

\nibf{Results.}
\input{secs/Tables/e2e_results}
\autoref{tab:e2e_results} shows the summarized performance results from the end-to-end evaluation across various metrics. 
We confirm that all nine trials of end-to-end evaluation successfully pinpoint the physical location of the target UE with 70\% distance error of 1.7m. 
Furthermore, we confirm that each \umaplus attack, where the four key steps are conducted in sequence, takes less than 5 minutes to determine the physical location of a single target UE from its online identity.
Not surprisingly, it takes about 2 minutes to perform the power boosting attack due to the process of acquiring radio configuration and manipulating subframes, as described in~\autoref{ss:power_design}. In multiangulation step, it also takes about 2 minutes to estimate the direction (\ie, angle) of the signal source at two sniffers and determine the location.

We further analyze the effectiveness of the power boosting attack on the end-to-end evaluation.
\autoref{tab:e2e_results} presents a performance comparison between two scenarios: the end-to-end evaluation with and without the power boosting attack.
The power boosting attack enhances localization performance, as evidenced by improvements in angular accuracy, localization accuracy, and arrival signal strength on the sniffers' side.
Additionally, in cases where we perform the end-to-end process without executing the power boosting attack, the physical location of the target remains undetermined at two out of nine locations, marked by gray circles in \autoref{fig:e2e_setup}. 

\input{secs/Figures/e2e_angle}

To investigate these differences, we measure the arrival signal strength according to the angle from the UE placed at the left gray circle in~\autoref{fig:e2e_setup}. 
We rotate the antenna from 0$^{\circ}$ (\ie, pointing to the UE) to $\pm$50\textdegree, using the sniffer placed at the bottom right.
As depicted in~\autoref{fig:e2e_angle}, in the absence of the power boosting attack step, the angle corresponding to the peak signal strength is not discernible. 
This directly leads to errors in determining the angle of the signal source, causing the attacker to pinpoint the physical location inaccurately.
This observation aligns with the discussion in \autoref{ss:challenge}, where it is noted that when the UE is in close proximity to the eNB, it generates its uplink signal with low transmission power. Additionally, \autoref{fig:e2e_angle} hints that power boosting can improve the direction finding capability by reducing the noise on the sniffer side when the attacker attempts to estimate the direction of signal source. Actually, the signal-to-noise ratio (SNR) in the UE direction has been found to be 30dB with power boosting and 13dB without power boosting.

\subsection{Power Boosting vs. Cellular Repeater}
\label{ss:strategy_eval}
We now examine the effectiveness of our strategy employing the power boosting attack to distinguish the UE's signal from the repeater's. 
Due to ethical concerns, the evaluation is conducted on the commercial testbed. 
The repeater's external antenna was installed near eNB and the UE was positioned 2m away from the repeater. We then placed the sniffer 5m away from the UE and the repeater's external antenna.

We first measure the arrival signal strength at the direction of the UE and the repeater's external antenna~(\autoref{tab:repeater}) without the power boosting attack. 
The arrival signal strength at the direction of the UE is measured as 5dBm. At the direction of the repeater, the signal strength is equivalent to 17dBm. 
Next, we measure the arrival signal strength in two directions, as before, while conducting the power boosting attack. 
Signal strength at the direction of UE, at this time, significantly increases to 25dBm, while the arrival signal strength at the direction of the repeater remains unchanged (static). 
As a result, by observing an increase in arrival signal strength after performing the power boosting attack, the \umaplus attacker can identify the UE's actual signal.

\input{secs/Tables/repeater}

%% file: secs/Tables/exp-summary.tex
\begin{table}[]
\centering
\caption{Test environments for each experiment.}
\vspace{-2mm}
\label{tab:exp-summary}
\resizebox{0.85\columnwidth}{!}{%
\begin{threeparttable}
\begin{tabular}{@{}p{6cm}l@{}}
\toprule
\textbf{Experiment}     & \textbf{Environment}                                \\ \midrule
RNTI acquisition        & ON       (\autoref{ss:rnti_eval})                                                \\
Scheduling manipulation attack & LT (\autoref{ss:sched_eval}), CT (\autoref{ss:integration})               \\
Power boosting attack   & LT (\autoref{ss:power_eval}), CT (\autoref{ss:integration})             \\
End-to-end evaluation of \umaplus    & LT (\autoref{ss:e2e_multi})                                                      \\
Defeating cellular repeater   & CT (\autoref{ss:strategy_eval})                                           \\
\midrule
Multiangulation using uplink signals & ON  (\cite{uma-site})                                     \\
Existence of shadow area  & ON  (\cite{uma-site})                         \\ 
Received strengths for LTE services  & ON  (\cite{uma-site})                         \\ \bottomrule
\end{tabular}%
\begin{tablenotes}
\scriptsize
\item ON: Operational network, LT: Lab testbed using srsRAN, CT: Commercial testbed.
\end{tablenotes}
\end{threeparttable}
}
\vspace{-3mm}
\end{table}

%% file: secs/Tables/tested_devices.tex
\begin{table}[!t]
\caption{COTS UE used in the experiments.}
\vspace{-3mm}
\label{tab:devices}
\centering
\resizebox{\columnwidth}{!}{%
\begin{threeparttable}
\begin{tabular}{llcc} 
\toprule
\multicolumn{1}{c}{Device} & \multicolumn{1}{c}{Baseband} & Lab testbed & Commercial testbed  \\ 
\midrule
Galaxy Note FE             & Exynos 8890                  & S+P & -           \\
Galaxy Note 10 5G          & Exynos 9825                  & S+P & -           \\
Galaxy S10 5G              & Exynos 9820                  & S+P & S+P         \\
Galaxy S20                 & Snapdragon 865               & S+P   & S+P         \\
LG G8                      & Snapdragon 855               & S+P   & -           \\
Huawei P30 Pro$^{\ast}$             & Kirin 980                    & S   & -           \\
Galaxy Z Flip4             & Snapdragon SM8475            & S+P   & S+P         \\
iPhone XS$^{\ast}$                  & Intel PMB 9955               & S   & -           \\
Redmi Note 9T 5G$^{\ast}$           & Mediatek Dimensity 800U      & S   & -           \\
\bottomrule
\end{tabular}
\begin{tablenotes}
\scriptsize
\item S: Scheduling manipulation attack~(\autoref{ss:sched_eval} and  \autoref{ss:integration}), P: Power boosting attack~(\autoref{ss:power_eval} and  \autoref{ss:integration}).
\item -- Only three types of devices were tested on the commercial testbed, due to compatibility issues with the environment.
\item $\ast$ Power boosting attack could not be confirmed on these devices due to the incompatibility of their baseband chipsets with our XCAL~\cite{xcal}, which is required to monitor uplink transmission power. However, given that their baseband chipsets are also designed to follow 3GPP standards, it is expected that power boosting attack should work on these devices.
\end{tablenotes}
\end{threeparttable}
}
\vspace{-5mm}
\end{table}

%% file: secs/Figures/e2e_setup.tex
\begin{figure}[!t]
\centering
\includegraphics[width=.95\columnwidth]{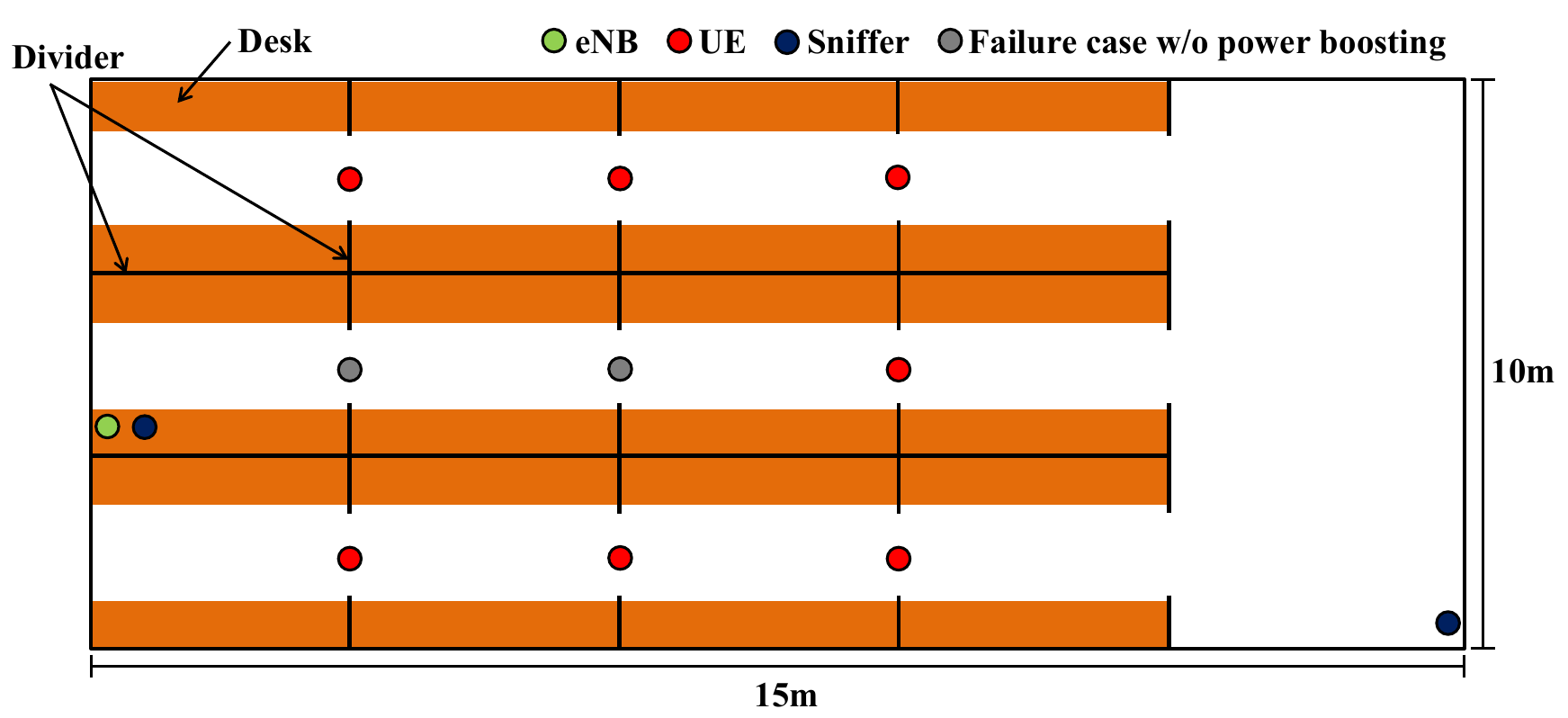}
\vspace{-3mm}
\caption{Experimental setup for e2e evaluation.}
\label{fig:e2e_setup}
\vspace{-4mm}
\end{figure}

%% file: secs/Tables/e2e_results.tex
\begin{table}[]
\centering
\caption{Evaluation of end-to-end experiment.}
\vspace{-2mm}
\label{tab:e2e_results}
\resizebox{\columnwidth}{!}{%
\begin{tabular}{@{}lcl@{}}
\toprule
\multicolumn{3}{c}{\textbf{Localization performance}}                                                     \\ \midrule
\multicolumn{1}{c}{Evaluation metrics }                              & \multicolumn{1}{c}{w/o power boosting attack} & \multicolumn{1}{c}{w/ power boosting attack} \\ \cmidrule{1-1}\cmidrule(lr){2-2}\cmidrule(lr){3-3}
Success rate                   & \multicolumn{1}{c}{77\%}  & \multicolumn{1}{c}{100\%}                                        \\
70\% angluar accuracy       & \multicolumn{1}{c}{$\sim$ 16$^{\circ}$}    & \multicolumn{1}{c}{$\sim$ 12$^{\circ}$}                                           \\
70\% distance accuracy      & \multicolumn{1}{c}{$\sim$ 2.8m}  & \multicolumn{1}{c}{$\sim$ 1.7m}                                         \\
Arrival signal strength (Max.)       & \multicolumn{1}{c}{-2 $\sim$ 19 dBm}      & \multicolumn{1}{c}{33 $\sim$ 38 dBm}                                        \\ \midrule
\multicolumn{3}{c}{\textbf{Time consumption for each step}}                                                                         \\ \midrule
RNTI acquisition                  & \multicolumn{2}{c}{$\leq$ 30 seconds}                                     \\
Scheduling manipulation attack & \multicolumn{2}{c}{$\leq$  30 seconds}                                     \\
Power boosting attack          & \multicolumn{2}{c}{$\leq$  2 minutes}                                      \\
Multiangulation                & \multicolumn{2}{c}{$\leq$  2 minutes}                                      \\ \bottomrule
\end{tabular}
}
\vspace{-4mm}
\end{table}

%% file: secs/Figures/e2e_angle.tex
\begin{figure}[!t]
\centering
\includegraphics[width=0.9\columnwidth]{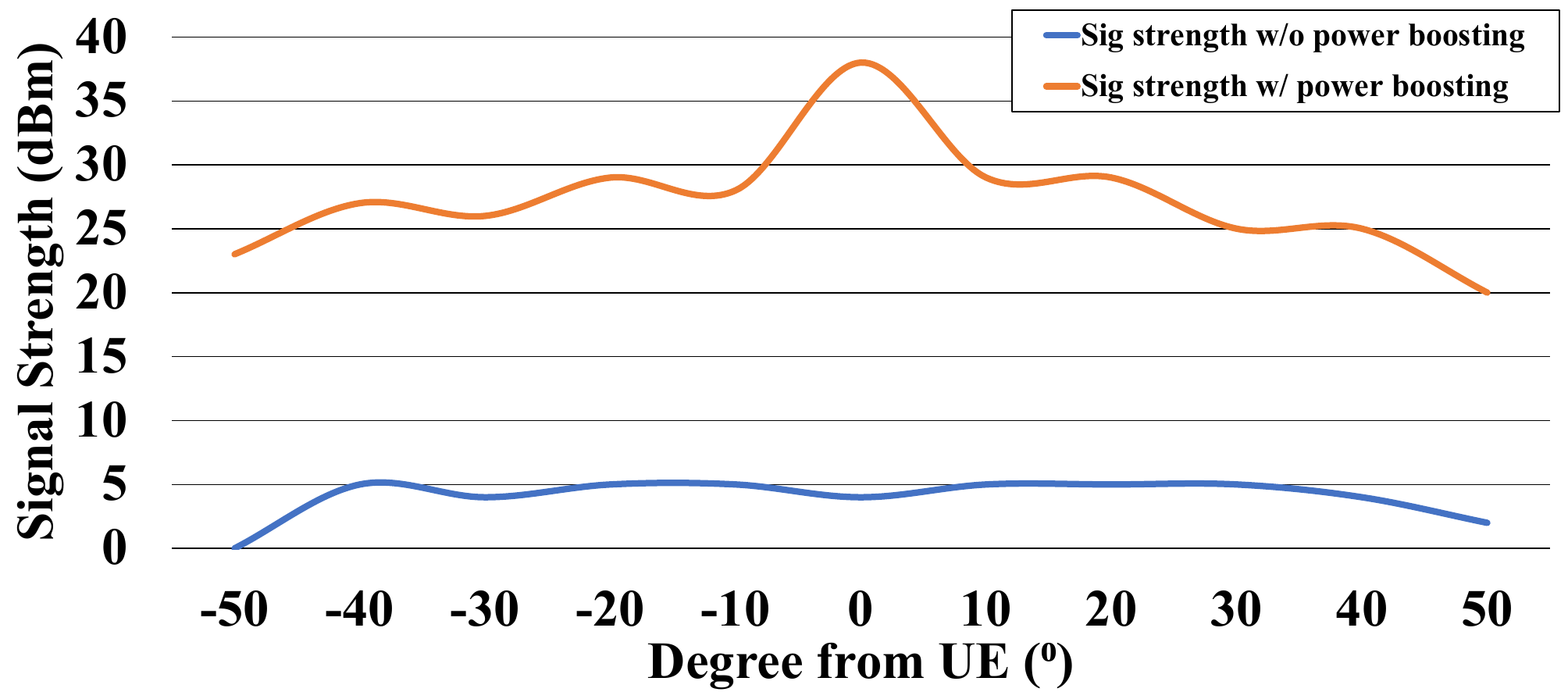}
\vspace{-3mm}
\caption{Impact of power boosting on DF-ing.}
\label{fig:e2e_angle}
\vspace{-5mm}
\end{figure}

%% file: secs/Tables/repeater.tex
\begin{table}[]
\caption{Impact of cellular repeater and TPC injection.}
\vspace{-2mm}
\label{tab:repeater}
\centering
\resizebox{0.8\columnwidth}{!}{%
\centering
\begin{threeparttable}
\begin{tabular}{lcccc} 
\toprule
\textbf{TPC injection}         & \multicolumn{2}{c}{Off} & \multicolumn{2}{c}{On}  \\ 
\cmidrule{1-1}\cmidrule(lr){2-3}\cmidrule(lr){4-5}
\textbf{Target}                & UE  & Repeater          & UE  & Repeater          \\ 
\midrule
\textbf{Signal strength (dBm)} & 5 & 17               & 25  & 17               \\ 
\bottomrule
\end{tabular}
\begin{tablenotes}
\scriptsize{
 \item [] Distance between the repeater's antenna and the UE is 2m.
}
\end{tablenotes}
\end{threeparttable}
}
\vspace{-5mm}
\end{table}

%% file: secs/10.5_Feasibility.tex
\subsection{Attack Feasibility}
\label{s:feasibility}
\nibf{Validity of experimental setup.} 
Due to legal and ethical constraints, we were unable to conduct end-to-end evaluations of \umaplus, including scheduling manipulation and power boosting attacks, in operational networks. These attacks could affect operational networks and normal user devices since they require signal injection over both downlink and uplink channels.

To address this limitation, we designed our experimental setup to mirror real-world settings as closely as possible. To demonstrate the attack feasibility, we used combinations of COTS devices and eNBs for various experiments, as shown in~\autoref{ss:real_eval}: 1) Nine COTS devices with different baseband chipsets manufactured by five vendors covering over 90\% of the market share~\cite{baseband_marketshare}, and 2) A commercial testbed featuring an eNB from Nokia, a leading eNB vendor, and a lab testbed built with open-source srsRAN~\cite{srsran} supporting a full-stack eNB.

\nibf{Standard-compliant feature exploitation.}
Our approaches exploit only standard-compliant features in LTE, including scheduling request, BSR, and TPC. Therefore, they should be effective on any standard-compliant eNBs and UEs (\ie, universal cellular environments). Additionally, GSMA stated, \textit{``We acknowledge that the mechanisms of power boosting attack and scheduling manipulation attack can make localization attacks more reliable,''} upon our responsible disclosure. This remark confirms that the observed result aligns with the expected standard-compliant behavior.

\nibf{Power boosting with SigOver.}
Power boosting leverages SigOver~\cite{yang2019hiding}, a well-known signal overshadowing attack shown to be effective in commercial networks~\cite{erni2022adaptover,yang2019hiding,tan2021data,kotuliak2022ltrack}. It injects malicious signals that overshadow legitimate ones, making the UE receive only the injected malicious signals due to capture effect and synchronization with targeted legitimate signals. Exploiting the capture effect and synchronization method of SigOver, a \umaplus attacker can increase the transmission power of the victim UE to the maximum.

%% file: secs/12_Discussion.tex
\section{Discussion}
\label{s:discussion}
\nibf{Other potential attacks using scheduling manipulation.}
Scheduling manipulation induces continuous uplink traffic generation on the target UE, potentially accelerating battery drain. To quantify this impact, we measured electric current levels in the UE's circuitry under both attack and normal conditions using a battery monitoring application~\cite{battery_monitoring}. Electric current directly correlates with battery consumption. During the scheduling manipulation attack, the average electric current level was measured at 340mAh, compared to 250mAh during normal cellular connection. This significant increase in the current level demonstrates that the scheduling manipulation attack substantially accelerates UE battery drain by forcing continuous uplink traffic generation.

We also analyzed the feasibility of the overcharging attack. Our analysis confirmed that scheduling manipulation cannot affect billing (\ie, charging) since the target UE's traffic generated by the attack consists only of padding, not actual data payload. In cellular networks, charges are incurred, when the Service Data Unit (SDU), the data payload for cellular service usage, is included in the packet and forwarded to the Packet Data Network Gateway (P-GW), specifically the Policy and Charging Enforcement Function (PCEF) in the core network, after decoding and decryption by the eNB. However, packets generated by scheduling manipulation don't include an SDU (refer to our website~\cite{uma-site}). Moreover, the attacker's packets aren't forwarded to the P-GW, as the attacker lacks the target's key for proper packet encryption.

Lastly, scheduling manipulation could potentially lead to the uplink resource draining attack, as demonstrated in~\cite{tan2021data}, where spoofing BSR results in unfair resource assignment.

\nibf{Limitations.}
First, the entire \umaplus was not tested end-to-end in operational cellular networks due to ethical concerns. Practical issues during physical localization may arise, such as multipath effects, especially in urban environments in reality. Fortunately, existing research has explored strategies for radio direction detection~\cite{treichler1983new,vanderveen1998estimation} and indoor localization~\cite{chintalapudi2010indoor, otsason2005accurate, zafari2019survey, kumar2014lte}. We believe integrating those algorithms can enhance localization accuracy in such environments. Second, while \umaplus should work in 5G networks based on specifications, it requires developing the necessary toolkit: a) 5G sniffer: LTESniffer~\cite{hoang2023ltesniffer} supports high-performance downlink and uplink sniffing in LTE, but existing 5G sniffer~\cite{ludant20225g} lacks uplink channel monitoring capabilities; b) 5G signal overshadowing: This has not been implemented yet; c) New 5G features: Wider bandwidth, beamforming, and massive MIMO may affect \umaplus design for 5G. Despite the requirements, attackers can still conduct \umaplus by downgrading the UE's connection through bidding down attacks. Finally, \umaplus could potentially be detected by the victim's device~(\autoref{ss:short_mitigation}), allowing them to turn off their device to avoid tracking. However, unless eNB/UE vendors implement attack detection, victims would need specialized equipment such as a sniffer, diagnostic monitoring tool (\eg, XCAL), and algorithms to differentiate attack packets from legitimate ones.

\nibf{Other application.}
\umaplus can be used in rescue operations for individuals inevitably unable to cooperate with localization. People who are injured, kidnapped, or cognitively impaired, may be unable to request rescue or report their location. 

\section{Countermeasure}
\label{s:countermeasure}
\subsection{Mitigation for Scheduling Manipulation}
\label{ss:short_mitigation}
\nibf{Mitigation at baseband.}
When the UE receives DCI 0 messages containing uplink resource allocation, it transmits packets over the uplink channel even if it did not request such allocation. 
To mitigate this issue, baseband vendors can implement a condition where data is transmitted only if the device previously requested uplink grants or transmitted a scheduling request. Alternatively, the device simply doesn't transmit data if there is no pending data in its buffer.

\nibf{Detection at base station.}
During scheduling manipulation, two anomalous behaviors occur; 1) The target UE reports it has no data to send in the current buffer, yet it repeatedly transmits packets over the uplink channel. 2) The uplink traffic contains only padding data as payload. These behaviors could potentially be detected at the eNB as anomalous.

\nibf{Reallocation of radio config.}
The eNB can enhance security by reallocating user-dedicated radio configurations over a secure channel after completing the security mode procedure. This reallocation can be done through \tts{RRC Connection Reconfiguration} message. By using this encrypted message to reassign the radio configuration, it becomes more challenging for an attacker to identify the target UE's specific radio configuration which is crucial for scheduling manipulation.

\subsection{Long-Term Mitigation}
\label{ss:long_mitigation}
Long-term mitigation requires modification of specification. Fundamentally, \umaplus exploits three vulnerabilities in lower-layer protocols of current cellular networks; 1) Lack of confidentiality in DCI messages, allowing location tracking, 2) lack of integrity protection in DCI messages, enabling the power boosting attack, and 3) scheduling manipulation attack due to lack of integrity protection in the uplink scheduling process. As 3) could be defended with short-term mitigation~(\autoref{ss:short_mitigation}), we focus on 1) and 2) here. Specifically, we discuss confidentiality and integrity protection for DCI 0, which manages both uplink resource allocation and transmission power.

Security protections for lower-layer cellular protocols have been explored~\cite{tan2021data,yang2019hiding}, but significant challenges remain in protecting DCI messages; 1) DCI is a PHY layer message, typically neither encrypted nor integrity protected by design, 2) only restricted space is reserved for DCI 0 messages (maximum 37 bits: up to 13 bits for resource allocation and 2 bits for TPC) to save radio resources, making it difficult to add message authentication code or digital signatures for integrity protection, 3) DCI is fundamental to cellular network operations, managing radio resources and controlling various communication aspects. Thus, protecting DCI messages would likely affect a significant portion of cellular technology design. Considering these challenges, we leave developing specific protection mechanisms for future work.

%% file: secs/11_Related.tex
\section{Related Work}
\label{s:related}
\input{secs/Tables/related}

\nibf{Physical localization.}
Kotuliak \etal~\cite{kotuliak2022ltrack} proposed a localization attack based on ToA measurement. Their approach exploits unencrypted information for the transmission time alignment (\ie, TA) and the actual monitored uplink arrival time. This ToA-based localization requires an additional process for fingerprinting the target device to account for hardware errors. 
The closest work to ours in the context of localization method (\ie, AoA) is LTEye~\cite{kumar2014lte}. It introduced indoor localization, utilizing a SAR to handle multipath issues.

Compared to \umaplus, existing models on physical localization didn't consider realistic scenarios as described in~\autoref{tab:comparison}. First, although they also provided RNTI acquisition for physical localization, LTrack~\cite{kotuliak2022ltrack} leverages the target's IMSI/TMSI, which is more difficult to obtain compared to online identity in realistic scenarios, and LTEye~\cite{kumar2014lte} necessitates an RF fingerprinting procedure with a corresponding database to trace the RNTIs. Second, they have limited opportunities for physical localization. They can trace the target UE only while it is in active status, connected to the eNB. This duration can be only 15--30 seconds, RNTI duration, in the worst case, restricting the attacker's localization capabilities. Additionally, even if the target UE is connected to the eNB, it may not transmit uplink traffic if there is no data to send in its current buffer, making the attacker unable to trace the target UE. Finally, they cannot correctly measure ToA or AoA when the target UE transmits its signals with low power or repeaters relay the target's signals. This restricts the attacker's signal detection range and makes physical localization difficult.

To overcome these challenges in the physical localization of uncooperative cellular devices, they should adopt our approach. Thanks to \umaplus, attackers can effectively achieve reliable fine-grained localization in realistic scenarios.

Through scheduling manipulation, \umaplus enables LTEye to perform physical localization without its practical requirements; 1) the target UE should be in an active status while the SAR's mobile antenna rotates 360\textdegree and 2) RNTI mapping process with RF fingerprinting is needed when the target's RNTI is reassigned. While we utilized multiangulation-based localization, \umaplus can also benefit the ToA-based multilateration technique. For instance, we can improve LTrack~\cite{kotuliak2022ltrack} utilizing multilateration, by addressing challenges such as tracking RNTI associated with online identity instead of IMSI extraction, ensuring continuous uplink transmission during multiple ToA measurements. Lastly, when LTEye and LTrack attackers face challenges stemming from low signal power or the presence of the repeater, they can perform the power boosting attack to improve their localization performance.

\nibf{Other cellular localization.}
Lakshmanan \etal~\cite{lakshmanan2021stealthy} demonstrated unprivileged location tracking by exploiting Carrier Aggregation (CA) side channels. However, it requires collecting CA measurements across all possible target paths, limiting the capability of localizers when attempting to trace cellular devices. Michalevsky \etal~\cite{michalevsky2015powerspy} showed that the location path is leaked by analyzing the power consumption of the target's device, but it requires installing malware on the device for location tracking. Previous studies~\cite{bitsikas2023freaky, baektargeted, cheng2023watching, hong2018guti,kune2012location,ludant20225g,shaik2016practical} aimed to determine the presence of the target UE within or beyond a specific cell coverage area by utilizing calls/SMSes. However, these approaches have restricted localization granularity, at best achieving cell-level accuracy.

\nibf{Message forgery attacks.}
Tan \etal~\cite{tan2021data} proposed data-plane signaling forgeries at unprotected layers in cellular networks. They exploited various vulnerabilities in lower-layer protocols such as BSR and TA to demonstrate threats derived from unprotected control messages. Studies~\cite{erni2022adaptover,yang2019hiding,tan2021data,kotuliak2022ltrack} have shown that unprotected cellular messages could be overwritten at the PHY layer through SigOver attacks.

%% file: secs/Tables/related.tex
\begin{table}[]
\caption{Comparison with existing models.}
\vspace{-2mm}
\label{tab:comparison}
\resizebox{\columnwidth}{!}{%
\begin{tabular}{l|c|c}
\hline
\multicolumn{1}{c|}{\textbf{Steps}}  & \begin{tabular}[c]{@{}c@{}}\textbf{Existing Physical}\\ \textbf{localization~\cite{kotuliak2022ltrack,kumar2014lte}}\end{tabular}                                    & \textbf{UMA}                                                                                       \\ \hline
1) RNTI acquisition         & \begin{tabular}[c]{@{}c@{}}From IMSI~\cite{kotuliak2022ltrack}, \\ RF fingerprinting~\cite{kumar2014lte}\end{tabular}                                     & \begin{tabular}[c]{@{}c@{}}From online identity \\ (\cite{hong2018guti, hussain2019privacy, kohls2019lost, kune2012location, ludant20225g, rupprecht2019breaking, shaik2016practical}) \end{tabular}                                                                              \\ \hline
2) Location tracking        & \begin{tabular}[c]{@{}c@{}}ToA~\cite{kotuliak2022ltrack}, AoA~\cite{kumar2014lte}\\ measurements\end{tabular}                                             & AoA measurements$^{\ast}$                                                                                   \\ \hline
2-1) Fixing RNTI$^{\dagger}$             & \begin{tabular}[c]{@{}c@{}}\textbf{C1:} Limited tracking time\\ (RNTI duration 15$\sim$30 sec)\end{tabular} & Unlimited tracking time                                                                            \\ \hline
2-2) Forcing uplink traffic$^{\dagger}$  & \begin{tabular}[c]{@{}c@{}}\textbf{C1:} Untraceable during \\ no traffic\end{tabular}                       & \begin{tabular}[c]{@{}c@{}}Traceable, regardless of \\ UE's status\end{tabular}                    \\ \hline
2-3) Power boosting         & \textbf{C2:} Limited detection range                                                                        & \begin{tabular}[c]{@{}c@{}}Improved signal detection \\ range (70m$\uparrow$)$^{\ddagger}$  and SNR\end{tabular} \\ \hline
3) Repeaters                & \begin{tabular}[c]{@{}c@{}}\textbf{C3:} Tracing repeaters, \\ instead of UE\end{tabular}                    & \begin{tabular}[c]{@{}c@{}}Distinguishing UE and \\ repeater \end{tabular}                                                                     \\ \hline
\end{tabular}%
}
\begin{tablenotes}
\scriptsize
\item $\ast$ Applicable to ToA measurements.
\item $\dagger$ Scheduling manipulation.
\item $\ddagger$ Refer to the shadow area graph on our website~\cite{uma-site}.
\end{tablenotes}
\vspace{-5mm}
\end{table}

%% file: secs/13_Conclusion.tex
\section{Conclusion}
In this paper, we investigate realistic challenges for the physical localization of uncooperative cellular devices down to the front door. We present \umaplus as a solution, enabling a localizer with only the target's online identity (\eg, phone number) to determine the target's physical location by actively exploiting three vulnerabilities in the LTE specification. While ensuring strict compliance with legal boundaries, we demonstrate the feasibility and effectiveness of \umaplus. We emphasize that \umaplus is universally applicable in any LTE network as it utilizes vulnerabilities inherent in cellular specifications. We have responsibly disclosed our findings to GSMA, who confirmed that \umaplus makes physical localization reliable. Finally, the countermeasures discussed in this paper could be considered for 5G and 6G implementations and specifications.

\begin{acks}
We sincerely appreciate reviewers and shepherd for their valuable comments. We would also like to thank Mincheol Son, Kwangmin Kim, Jiwoo Seo, Hansung Bae, Taisic Yun, and Beomseok Oh for their assistance in improving our paper. This research was supported and funded by the Korean National Police Agency [Tracking and identifying devices and call traffic in voice phishing ecosystem / PR10-03-020-22].
\end{acks}